\documentclass{bmcart}

\usepackage{lineno,hyperref,array,pbox}
\usepackage{graphicx}
\usepackage{subcaption}
\modulolinenumbers[5]
\usepackage[utf8]{inputenc} %unicode support
\usepackage[strings]{underscore}

\begin{document}

\begin{frontmatter}

\begin{fmbox}
\dochead{Research}

\title{Situating Agent-Based Modelling in Population Health Research}

\author[
   addressref={aff1},                   % id's of addresses, e.g. {aff1,aff2}
   corref={aff1},                       % id of corresponding address, if any
   %noteref={n1},                        % id's of article notes, if any
   email={eric.silverman@glasgow.ac.uk}   % email address
]{\inits{ES}\fnm{Eric} \snm{Silverman}}
\author[
   addressref={aff1},
   email={umberto.gostoli@glasgow.ac.uk}
]{\inits{UG}\fnm{Umberto} \snm{Gostoli}}
\author[
   addressref={aff1},
   email={stefano.picascia@glasgow.ac.uk}
]{\inits{SP}\fnm{Stefano} \snm{Picascia}}
\author[
   addressref={aff1},
   email={jonatan.almagor@glasgow.ac.uk}
]{\inits{JA}\fnm{Jonatan} \snm{Almagor}}
\author[
   addressref={aff1},
   email={mark.mccann@glasgow.ac.uk}
]{\inits{MM}\fnm{Mark} \snm{McCann}}
\author[
   addressref={aff3},
   email={richard.shaw@glasgow.ac.uk}
]{\inits{RS}\fnm{Richard} \snm{Shaw}}
\author[
   addressref={aff2},
   email={claudio.angione@tees.ac.uk}
]{\inits{CA}\fnm{Claudio} \snm{Angione}}

%%%%%%%%%%%%%%%%%%%%%%%%%%%%%%%%%%%%%%%%%%%%%%
%%                                          %%
%% Enter the authors' addresses here        %%
%%                                          %%
%% Repeat \address commands as much as      %%
%% required.                                %%
%%                                          %%
%%%%%%%%%%%%%%%%%%%%%%%%%%%%%%%%%%%%%%%%%%%%%%

\address[id=aff1]{%                           % unique id
  \orgname{MRC/CSO Social and Public Health Sciences Unit, University of Glasgow}, % university, etc
  \street{200 Renfield Street},                     %
  \postcode{G2 3AX}                                % post or zip code
  \city{Glasgow},                              % city
  \cny{UK}                                    % country
}
\address[id=aff2]{%
  \orgname{Department of Computer Science and Information Systems, Teesside University},
  %\street{D\"{u}sternbrooker Weg 20},
  \postcode{TS1 3BX}
  \city{Middlesbrough},
  \cny{UK}
}
\address[id=aff3]{%                           % unique id
  \orgname{Institute of Health and Wellbeing, University of Glasgow}, % university, etc
  %\street{200 Renfield Street},                     %
  \postcode{G12 8RZ}                                % post or zip code
  \city{Glasgow},                              % city
  \cny{UK}                                    % country
}

\end{fmbox}

\begin{abstractbox}

\begin{abstract} % abstract
%Research in population health has been highly successful, as evidenced by its contribution to significant health improvements throughout the developed world.
Today's most troublesome population health challenges are often driven by social and environmental determinants, which are difficult to model using traditional epidemiological methods.  We agree with those who have argued for the wider adoption of agent-based modelling (ABM) in taking on these challenges.  However, while ABM has been used occasionally in population health, we argue that for ABM to be most effective in the field it should be used as a means for answering questions normally inaccessible to the traditional epidemiological toolkit.  
%unsuitably compared on a like-for-like basis with conventional epidemiological approaches, when the methodology is better-suited to answering different research questions.  
In an effort to clearly illustrate the utility of ABM for population health research, and to clear up persistent misunderstandings regarding the method's conceptual underpinnings, we offer a detailed presentation of the core concepts of complex systems theory, and summarise why simulations are essential to the study of complex systems.  We then examine the current state of the art in ABM for population health, and propose they are well-suited for the study of the `wicked' problems in population health, and could make significant contributions to theory and intervention development in these areas.
\end{abstract}

%%%%%%%%%%%%%%%%%%%%%%%%%%%%%%%%%%%%%%%%%%%%%%
%%                                          %%
%% The keywords begin here                  %%
%%                                          %%
%% Put each keyword in separate \kwd{}.     %%
%%                                          %%
%%%%%%%%%%%%%%%%%%%%%%%%%%%%%%%%%%%%%%%%%%%%%%


% MSC classifications codes, if any
%\begin{keyword}[class=AMS]
%\kwd[Primary ]{}
%\kwd{}
%\kwd[; secondary ]{}
%\end{keyword}

\end{abstractbox}

\end{frontmatter}

%\linenumbers

\section{Introduction}

%MMcC new draft 14/2/18 - Start

%There are few settings where a deduction from a model will directly determine a policy decision; these settings tend to be those with low uncertainty, low risk, and where competing views regarding the situation at hand are absent. None of these apply when our focus is the distribution of health in populations, and the factors that contribute to it. Taking a linear view of the progression from scientific evidence to applied decisions oversimplifies the multiple and competing interests, values, and outcomes for stakeholders which have an input into any action that will ultimately be taken \cite{oliver2014}.

Health policy-making is, at the root, an attempt to undertake principled decisions in an environment of high uncertainty and high risk, in which competing pressures and interests from stakeholders have a significant effect on the actions ultimately taken \cite{oliver2014}.  The rise of causal inference methods (CIM), has attempted to address this by providing a framework under which we may predict the outcomes of proposed interventions.  

However, CIM as applied in epidemiology today, has coalesced around a set of tools with certain limitations when applied to complex systems.  For example, directed acyclic graphs (DAGs) are frequently used in CIM, but DAGs are unsuitable for modelling systems containing feedback loops (given they are acyclic), a common feature of complex social systems.  Critics of CIM also suggest it takes an overly linear view of the decision-making process leading from evidence to decision-making, and have called for alternative concepts of cause that are not solely based on probabilistic statements about population outcomes in alternative worlds \cite{krieger2016}. Decisions over causality can be based on pragmatic pluralism \cite{vandenbroucke2016}, or inference to the best explanation \cite{lipton2003,krieger2016} (often characterised as a form of abductive reasoning).  Decision-makers ultimately must make decisions, even while accepting that their evidence is incomplete or flawed or both, and that the exact causal process underlying the system of interest is still uncertain.

Here we propose that the addition of Agent-Based Modelling (ABM) and related complex-systems-based approaches to the population health toolkit will enable better-informed and more robust decision-making in population health.  ABMs allow for the representation of causal processes in systems that include feedback loops and multiple layers of complex, interacting components.  ABMs can model explicitly the individual-level decision-making that can lead to unexpected \emph{emergent} effects at the population level.  The necessity of constructing informed conceptual models of system processes when building an ABM also opens up new avenues for co-production of models with decision-makers, while also facilitating the inclusion of both quantitative and qualitative data.  Taken in combination, these properties of ABMs enable them to serve as a complementary approach to the current population health toolset, and in turn to enable both scientists and policy makers to better navigate the non-linear abductive process leading from scientific evidence to policy action.

\section{The need for a new approach to decision-making}

%An effective intervention successfully disrupts a pathway between a cause and an effect. While there has been a long recognition of upstream social determinants of health, most historical successes in public health have focussed on modifying proximal causes, where a clear causal link has been postulated and upon which an intervention aims to exert influence. 

Research in population health has to its credit a number of major successes over the years, due to concerted action on communicable disease through improved sanitation and mass immunisation programmes, prompting major progress on serious health problems such as lung cancer and heart disease and a steady increase in life expectancy until recently \cite{hiam2018}.  More recently, significant progress has been made on noncommunicable diseases through tobacco control and lifestyle education.  However, `despite major investment in both research and policy, effective action to tackle pressing contemporary public health challenges remains elusive' \cite{rutter2017}. The most intractable health problems are those that spring fundamentally from more complex causes: behavioural and social influence; and environmental interaction.  Some notable `wicked' problems \cite{Andersson2014} include obesity, alcohol and drug misuse, and the persistence of health inequalities, among others.

The `wicked' health problems of the 21st century are driven by numerous influences where the causal links are not clearly defined, and the mechanisms that influence them are elusive \cite{rittel1973}. Wicked problems in health are serious challenges for policy makers, given that such problems `are continually evolving; have many causal levels; have no single solution that applies in all circumstances and solutions can only be classified as better or worse, rather than right or wrong' \cite{signal2013}.  Behavioural risk factors underpin these wicked health problems, but the evidence is sparse as to how physical and social environments influence health behaviours, and what can be done to improve them. 

Identifying, implementing and evaluating effective responses to major population health challenges requires a wider set of approaches beyond the traditional methods of public health research and should involve a wider set of actors beyond the health services \cite{academy2016}. Population health is influenced by multiple interacting determinants, including social, political, environmental, biological and behavioural factors. Current challenges to the health of the public, and the overarching inequalities in health within and across populations, are resistant to simple, linear, silver-bullet approaches \cite{rutter2017}.

Traditional epidemiological methods face the most difficulties in these types of population health challenges, in which determinants of illness and disease are multifaceted, inter-related and non-linear \cite{galea2010}.  Unravelling the complex interactions of social and environmental determinants is challenging when using statistical methods in which individuals and their actions may not be explicitly represented.  Individual-based modelling practices such as microsimulation can model some of these behavioural and spatial effects, but the effects of interactions between individuals and environmental factors are still hard to capture.

We claim that the study of `wicked' health problems necessitates the investigation of human society as a complex system, and that public health itself can be conceptualised as the systemic \emph{emergent outcome} of a complex system. Complex systems in this context can be defined as systems composed of interacting adaptive agents. More precisely, the dynamics of a complex system are driven by the interaction taking place at the level of its components, and the components' adaptations to the environmental changes they mutually generate. As such, social systems display some characteristic properties which do not lend themselves to reductionist approaches, but instead necessitate the adoption of `bottom-up' models, i.e. models that generate aggregate patterns starting from an explicit representation of the behaviour of the components of the system -- in this case, individuals; the direct interactions between the components; and interactions between the components and their environment. 

Consequently, we propose that complex systems simulation methods, in particular various forms of agent-based modelling (ABM), are a critical component in the fight against the `wicked' health problems facing us in the 21st century and should become a key part of the toolkit for tackling public health problems. These modelling techniques allow us to represent individual behaviours and their interactions, study the tangled web of causal relationships among environmental, physical and social factors affecting health-related behaviours, and simulate the effect of these relationships on the dynamics of public health problems at the population level.

Ultimately, we take the view that decision-making related to complex, `wicked' health problems must include the use of simulation in order to address these issues in a coherent way.

In the remainder of this article, we will summarise the key characteristics and properties of complex systems, explain the utility of ABMs for the study of complex problems in population health, and critique the previous methodological debate on ABMs in this area.

%In the remainder of this article we will summarise the current state of play in ABMs for health, critique their application and the methodological debate surrounding their use, describe issues that have been missed or overlooked among the ABMs for health community, and set out recommendations for the use of ABMs and the practice of research to improve the health of the public.

%We will detail the areas in which ABMs and related conceptual frameworks can supplement the work already being done in the field, and will provide examples of current and ongoing work which is developing ABM methods specifically tuned for population health applications.  In doing so we hope to provide inspiration and motivation for further work in this area, and develop the foundations of a complex systems science methodology for population health research.

\section{Challenges from complexity in public health}

In line with our claim that public health is an emergent property of a complex system, in this section we explain the key characteristics of complex systems, the ways in which human society fulfills these characteristics, and the resultant impact on research efforts related to human social systems. 
 
\subsection{Emergence}
A fundamental property of complex systems is \emph{emergence}, which philosopher Mark Bedau divided into strong and weak forms in the context of complex systems \cite{bedau97b}.  Frequently, references to emergence in the natural sciences and philosophy are referring to \emph{strong emergence}, which describes properties of systems which are not deducible from the behaviour of their component parts. To paraphrase Bedau's example, the inscrutable phenomenon of consciousness is clearly a consequence of neural activity, yet our knowledge of the behaviour of neurons does not provide us with any insight into the function of consciousness \cite{bedau97b}.  Consciousness is not a property exhibited by individual neurons, and appears distinct from any particular neural property or behaviour, and still it arises from neural activity.  Consciousness can also change our neural activity, despite being distinct from it; this is \emph{downward causation}, meaning that an emergent property can alter the behaviour of the component parts from which it emerges.   

%In the context of the development of complexity science and the subsequent popularity of ABMs, modellers and philosophers alike have devoted considerable time to the exploration of emergence and its consequences. 

Strong emergence, while successfully capturing the idea of a macro-level property which is distinct from and yet capable of influencing its own components -- as opposed to a macro-property that is merely an interesting consequence of micro-level activity -- is philosophically problematic.  Strongly emergent properties appear to be essentially autonomous from their components, and yet are able to exert strong causal influence on those same properties.   

Consequently, Bedau's exploration of \emph{weak emergence} has become an important concept for complex systems science:
\begin{quote}
    Macrostate P of S [system composed of micro-level components] with microdynamic [micro-level behaviour] D is weakly emergent iff [if and only if] P can be derived from D and S's external conditions, but only by simulation (Bedau 1997 p. 4)
\end{quote}

In other words, the behaviour of a system composed of interacting micro-level components is ultimately derived from its micro-level behaviours and the influence of its environment.  If we can simulate these interactions explicitly, we can simulate the dynamic that generates the emergent property, and thus we can replicate the emergent macrostate of that system.  Weakly-emergent properties are accordingly more accessible to scientific study, in that their emergence can be replicated via step-by-step simulation of the interaction of their constituting components and the surrounding environment.  Conversely, if we do not simulate this dynamic, we cannot replicate the weakly-emergent behaviour.  

\begin{figure}
    \centering
    \captionsetup{width=.8\linewidth}
    \begin{subfigure}[b]{0.4\textwidth}
        \includegraphics[width=\textwidth]{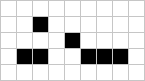}
        \caption{7-cell initial condition}
        \label{fig:life1}
    \end{subfigure}\hspace{0.12\textwidth}
    \quad
    \begin{subfigure}[b]{0.3\textwidth}
        \includegraphics[width=\textwidth]{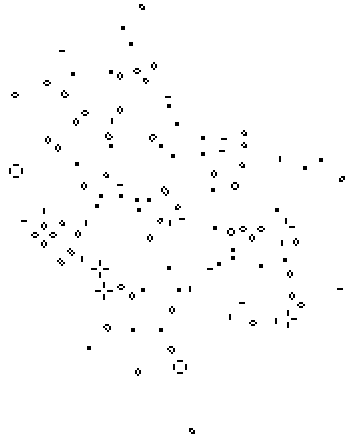}
        \caption{After 5,206 steps}
        \label{fig:life2}
    \end{subfigure}
    \caption{An example of the unexpected complexity of simple patterns in the Game of Life.  This 7-cell pattern is called an `acorn' and stabilises after 5,206 steps with a population of 633 live cells.}\label{fig:life}
\end{figure}

This latter point is particularly important, as the macro-level emergent outcomes of a system's microdynamic cannot be straightforwardly predicted, even with perfect knowledge of a system's initial state and the rules driving its microdynamic.  Bedau illustrates this using the Game of Life, a famous computational system in which cells on a grid change state according to the states of their neighbours.  Cells in Life have only two states -- alive or dead -- and their future states are determined by very simple rules according to how many of their neighbouring cells are alive or dead \footnote{Life's rule is often expressed simply as B3S23: cells are (B)orn when they have exactly three live neighbours, and (S)urvive if they have two or three live neighbours.  In all other circumstances, cells die or remain dead.}.  Despite this simplicity, even very simple starting configurations in Life can produce remarkably complex behaviour (see e.g. Figure \ref{fig:life}), and Life can even play host to patterns capable of replicating any possible computation (a property known as \emph{computational universality}) \cite{gardner83,rendell2002turing}.  As Bedau notes, this has profound implications:

\begin{quote}
    With few exceptions, it is impossible without simulation to derive the macrobehaviour of any state in a Life configuration even given complete knowledge of that configuration.  In fact, since a universal Turing machine can be embedded in Life, the undecidability of the halting problem proves that in principle there can be no algorithm for determining whether the behaviour exhibited in an arbitrary Life world will ever stabilize. Yet all Life phenomena can be derived from the initial conditions and the birth-death rule. (Bedau 1997 p. 14)
\end{quote}

Thus, the only way we can replicate the weakly-emergent macrostates of the Game of Life is to simulate its behaviour step-by-step.  By extension, given that most complex systems will have significantly more complicated microdynamics than the Game of Life, replicating the macrostates of weakly-emergent systems requires the use of simulation to replicate their microdynamics.      

% a property is defined as weakly emergent if that property is unexpected given the rules that govern the system's component parts (i.e. the system's micro-units), even if those rules are well-understood \cite{bedau97b}.  Bedau suggests that these properties are particularly amenable to being simulated; Chalmers illustrates this using the example of cellular automata, in which simple, easily understood calculation rules can produce unexpected higher-level patterns \cite{chalmers2006}. In this context, the term 'complex' has a relative meaning and also brings to light our cognitive limitations -- our inability to run mental simulations allowing us to forecast the system's dynamics, even when it runs based on relatively simple rules. In other words, due to our limited cognitive capabilities, we are unable to qualitatively forecast the aggregate outcome of the interplay between the system's micro-units \cite{berkman2011}.

\begin{figure}
    \centering
    \begin{subfigure}[b]{0.4\textwidth}
        \includegraphics[width=\textwidth]{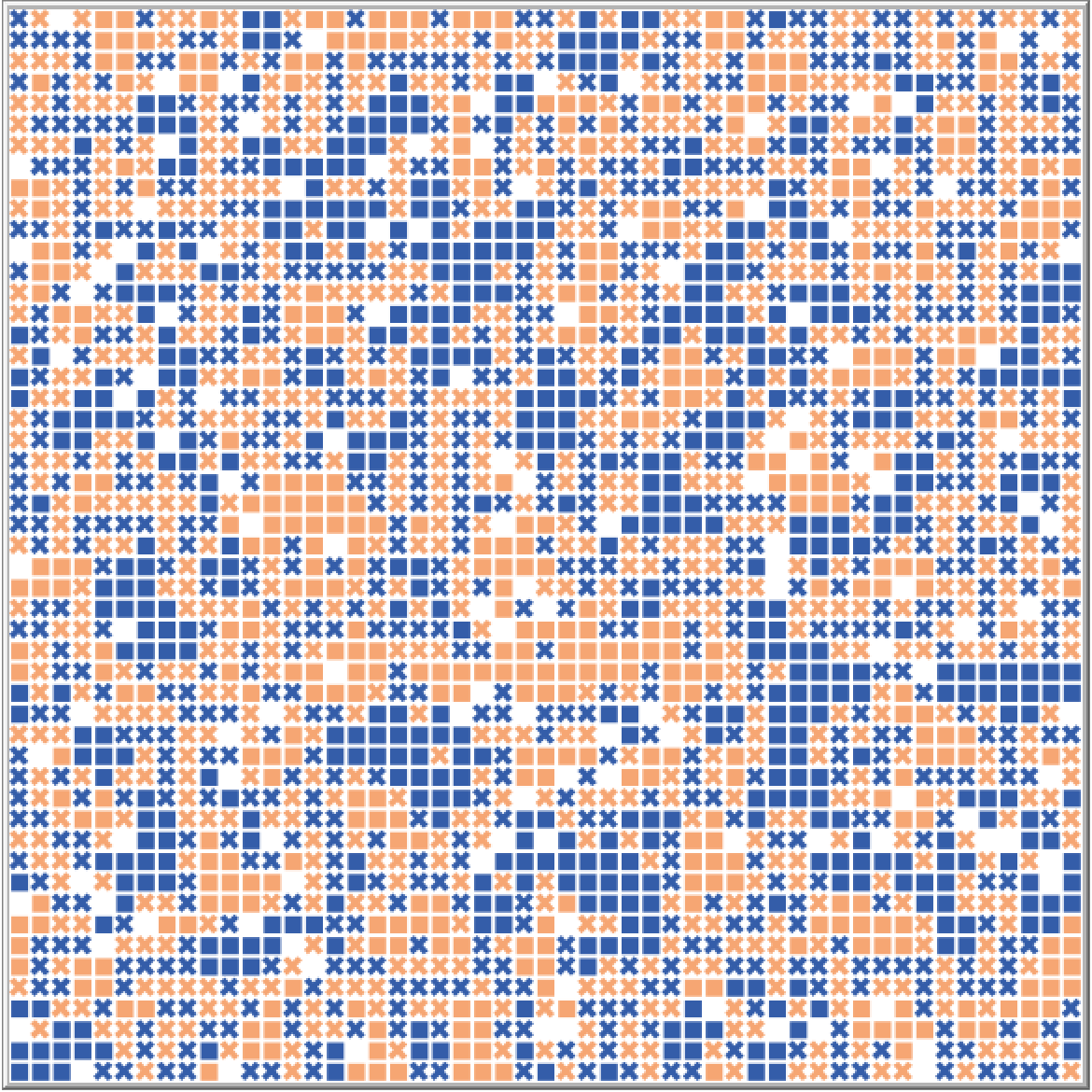}
        \caption{The start of the simulation}
        \label{fig:schelling1}
    \end{subfigure}
    \quad
    \begin{subfigure}[b]{0.4\textwidth}
        \includegraphics[width=\textwidth]{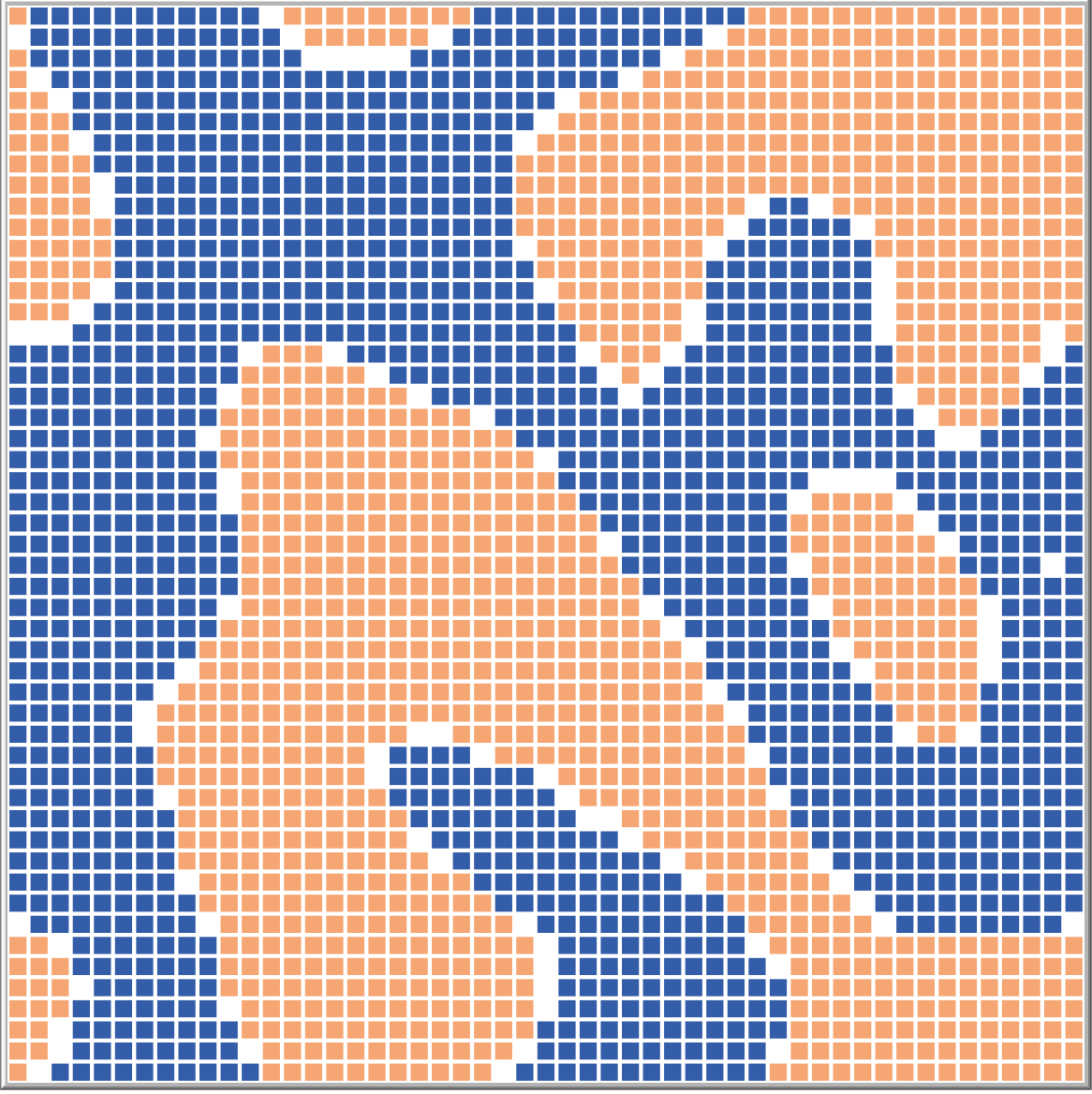}
        \caption{After 116 steps}
        \label{fig:schelling2}
    \end{subfigure}
    \caption{Sample run of the Schelling segregation model}\label{fig:schelling}
\end{figure}

A well-known example of an agent-based model replicating a weakly-emergent phenomenon from the interaction of micro-level entities is given by Schelling's residential segregation model \cite{schelling71, schelling78}. In this model, very simple agents are living in a virtual grid-based world, and at each discrete step of the simulation are able to choose to move their location.  Their decision to move is based on a preference for the group composition of their neighbourhood; if the number of their neighbours belonging to a different group than themselves is above a certain threshold, the agent will move to a new random square on the grid.  Here the agents' segregation is the emergent property of the system, while the preference for in-group neighbours is the parameter driving the agents' behaviour. Schelling showed that even a relatively low threshold generates a high degree of residential segregation, a result which is not predictable solely by knowing the agents' behavioural rules (see Figure \ref{fig:schelling}).  Thus we can describe Schelling's model as \emph{weakly emergent}, given that its macrostate is derivable only by simulating the system step-by-step, despite its known and very simple behavioural rules.

Similarly, the phenomena of interest in public health research -- from the development of health inequalities to the spread of obesity in certain communities -- are consequences of complex interactions between individuals and their physical and social environments. Successful interventions may seek to influence the phenomena by altering individual, low-level behaviours through a number of different routes, in the hope that the population-level picture which emerges from those actions changes for the better -- much like the simulation scientist tweaking the preferences of Schelling's agents in the hope of reducing segregation. Following Axelrod and Tesfatsion, we might align ourselves to using ABMs for normative understanding, or `evaluating whether designs proposed for social policies, institutions, or processes will result in socially desirable system performance over time' \cite{axelrod05}.

If we accept that population-level health patterns are weakly-emergent phenomena deriving from the interactions between individual, society and environment, then it follows that answering some questions about those properties will require simulating those interactions explicitly via simulation.  Agent-based modelling, as a methodology tailored to the investigation of emergent properties, is well-placed to provide insight into these phenomena by explicitly modelling both individual behaviours in response to an intervention and interactions between individuals and with their environment.  

\subsection{Non-linearity}
In the context of complex systems science, there are two extant definitions of `non-linearity'.  The first, which we will refer to as \emph{causal non-linearity}, encompasses the manner in which complex systems tend to be characterised by cyclic, relational and mutual causal relationships between the variables describing the system's state.  This characteristic of complex systems means that popular causal inference methods like directed acyclic graphs (DAGs) cannot be used to characterise complex systems, as DAGs cannot include feedback loops.

%In describing complex systems and their behaviour, we must distinguish between two different acceptations of 'non-linearity' of a complex system. The first kind of `non-linearity' refers to causality: in a system displaying causal linearity, causation is directed in one single direction. System dynamics has repeatedly shown that, for many real-world systems, linear causal models are the exception rather than the rule, as these systems are often characterized by cyclic, relational and mutual causal relationships among the variables describing the system's state. 

%\textit{Diffuse causality} is one variety of non-linear causality in which the behaviour of a micro-unit indirectly affects the behaviour of other agents via its effect on the agents' local environment -- a `chain reaction' through which the initial effect propagates through the system. The Schelling model again offers a clear example of this kind of non-linear causal relationship: an agent moving to another location changes the local environment of both the old and the new location's neighbors, affecting the probability that they will too move to another location. 

The Schelling model again provides a useful example of this concept.  Schelling's simple agents change the local environment of both their previous and current neighbours, thereby affecting the probability that those neighbours may move to another location.  In this way, the agents are affecting the actions of other agents indirectly via their shared environment. 

Applying this concept in the context of public health, we might imagine an ABM that examines how unhealthy behaviours propagate in a social network by explicitly representing agents with their characteristics, social relations and interactions.  For example, a simulation can explore the spread of unhealthy habits (such as smoking, drinking, and drug-use) among socially-related individuals, and evaluate the effectiveness of various strategies of network intervention that intend to induce desirable behavioural change across the social network.

Complex systems are characterised by a tangled web of non-linear causal relationships, which blurs the distinction between exogenous and endogenous variables. The non-linear causal relationship among the system's components and between the components and their environment is the fundamental reason behind our inability to forecast the system's dynamics from the components' behavioural rules: the web of causal relationships is simply too tangled for our limited cognitive capabilities to take all of them into account when trying to run a mental simulation of the system. 

The second definition of `non-linearity' refers to the kind of relationship between one or more exogenous variables and one endogenous variable: in this case, saying that the relationship is not linear means that variations in the exogenous and the endogenous variables are not proportional. In the Schelling model, we may change the agents' tolerance for different neighbours without noticing any significant change in residential segregation at the population level, as long as we are below or above a threshold level. Once we reach this level, however, a slight change in the agents' tolerance changes the system from a mixed state to a segregated one. Thus, the relationship between the agents' tolerance and the system's level of residential segregation is a non-linear one. 

A similar relationship is found in the dynamics of infectious disease. The spread of an infectious disease is defined by the basic reproduction number, which is the average number of new cases caused by an infected individual during his infectious period. The basic reproduction number is a threshold that dictates whether the infection will persist over time. When the reproduction number is lower than one, the infection can not persist in the population, whereas if it is greater than one the infection will spread and persist. Non-linearity is significant to both the Schelling model and in infectious disease modelling, since it demonstrates how past trends may abruptly change once a certain threshold is crossed, producing a qualitative change in the state of the system: from mixed to segregated and from a non-persistent infection to an epidemic. 

Of course, these two kinds of 'non-linearity' are strictly related. Complex systems are characterized by non-linear causal relationships between its components, and thus we often observe a non-linear relationship between exogenous and endogenous variables at the aggregate level.

\subsection{Adaptive behaviour}
Adaptive behaviour refers to the capacity or propensity for an agent to change its state following a change in its environment (including the behaviour of its neighbours)\footnote{We should note here that adaptive behaviour is an area of study in its own right within complex systems, with a particular focus on \emph{embodiment} and \emph{embodied cognition} -- the theory that the cognition and behaviour of living systems is shaped by the interaction between their physical form and the surrounding environment, not just the workings of the brain itself.  In this framework, a complete description of the behaviour of an organism requires study not just of the organism itself, but the entire \emph{brain-body-environment system}.  This area of work has a strong computational tradition and is closely tied to work in complex systems, dynamical systems and artificial life.  Here we use the term \emph{adaptive behaviour} in a more general sense.  For more on embodiment, the works of Francisco Varela et al. (\emph{The Embodied Mind: Cognitive Science and Human Experience}) and Andy Clark (\emph{Being There: Putting Brain, Body and World Together Again}) are excellent starting points.}. This fundamental characteristic of complex systems allows for non-linear causality: an environmental variation prompts the agents' behavioural responses, which then feed back into additional environmental variations, and so on. In other words, the system's components may affect each other both directly and indirectly through changes in their common environment. 

For example, consider the development of a new road that passes through a neighbourhood that will increase traffic, noise and air pollution in the area. As a result, residents who can afford to move may leave, and local housing prices may decline. The decline in housing prices may attract a less affluent population to the area. In this situation, individuals are adapting to new conditions in the environment, and the system -- and thus the neighbourhood -- self-organises as a result. In the new order that emerges an increasingly deprived population is located in a neighbourhood with poor environmental conditions and exposed to greater health risks.

Given that the components are the `engines' of the co-evolutionary process driving the system's dynamics, the behavioural model of these components is the fundamental building block of any complex systems model.  As these are weakly-emergent phenomena, we cannot replicate the dynamics of the system unless we simulate it as the result of the interaction between the system components and their environment.

Human societies are characterised by adaptive behaviour of the most complex kind, as human beings are able to recognise that they are in a complex system, identify the system's emergent properties and develop models that take them into account to drive their own actions. This phenomenon of \emph{second-order emergence}, or the fact that emergent social institutions become part of the agents' models driving their behaviour, create direct causal relationships between the components' behaviour and the system's dynamics, which further compounds the complexity of the system\cite{gilbert00}.

%As Gilbert put it:

%\begin{quote}
%In the physical world, macro-level phenomena, built up from the behaviour of micro-level components, generally themselves affect the components....
%The same is true in the social world, where an institution such as government self-evidently affects the lives of individuals. The complication in the social world is that individuals can recognise, reason about and react to the institutions that their actions have created. Understanding this feature of human society, variously known as second-order emergence, reflexivity and the double hermeneutic, is an area where computational modelling shows promise \cite[][p. 5]{gilbert00}.
%\end{quote}

\subsection{The complex systems challenge to traditional epidemiology}
Having outlined the defining characteristics of complex systems, we can better understand why they pose a challenge to the statistical approach typically adopted by epidemiology and how ABM can help epidemiology to rise to this challenge. Public health problems can be seen as the emergent outcomes of the complex social system that is human society. As such, to understand their dynamics we need to develop models based on the explicit representation of the components of society -- individual human beings. 

Individuals' adaptive behaviour and the resulting web of causal relationships between agents and their environment mean that non-linear relationships between system variables are pervasive in human social systems.  This means that a very small variation of system inputs can generate a big variation in system outcomes, or vice versa. We can visualise a non-linear complex system as one where the space of possible outputs is very rough along the many input dimensions: because of the number of factors affecting the relationship between any two variables, points that are very near to each other in the space of any input can be far apart in the space of outcomes. While the traditional statistical approach can be used \textit{in principle} to shed light on the causal mechanism through which variable X affects variable Y (and in fact much epidemiological research consists of the addition of confounders and mediators to the original theoretical model to enhance our understanding of how variable X affects variable Y), success relies on the availability of a `sufficient' number of observations for the analysis to have enough statistical power, a threshold that increases with the number of confounding variables in the causal model.  
This represents a limit to the complexity of the theoretical model that can be statistically analysed. 

With respect to complex systems like human society, this creates two major problems. First, a complex system may contain variables for which it is difficult or impossible to gather empirical values.  Second, even if our theoretical model does not contain such variables, in complex systems the number of \textit{potentially} conditioning variables is typically very large, so we may have too few observations to conduct a meaningful statistical analysis of the relationship between the variables of interest, or reach the limits of analytic tractability of a mathematical model with dozens of variables.

Thus, we see the reason why most causal models in traditional epidemiology are relatively simple compared to ABMs: the number of observations must be large enough for the analysis to have the desired statistical power, while remaining analytically solvable.  In other words, our tools force us to assume that numerous variables which we may ideally want to include in our models do not affect the relationship between X and Y.  We call this the \emph{stability assumption}, in that it requires that the relationship of interest is unaffected by changes in contextual variables.

%This is the reason why most of the causal models underlying the traditional epidemiological approach are relatively simple (simple enough to be based, in many cases, on an implicit causal mechanism): they need to be that way in order for the number of observations to be large enough for the analysis to have the desired level of statistical power, and remain analytically solvable. In other words, the traditional epidemiological approach is forced, by the tools it adopts, to assume that many variables, which we may want to include in our model \textit{in theory}, do not affect the relationship between X and Y, an assumption which we may call the \textit{stability} assumption, as it requires that the relationship of interest is not affected by changes in other, `contextual', variables.

Statistical approaches suffer further when data is sparse, as is often the case in human social systems.  Properly-specified theoretical models can still be applied in these cases as means for increasing our understanding of system behaviour; such models can form the basis for the examination of `what if' scenarios and for probing system behaviour via sensitivity analysis.  We propose that ABM can be very effective in this regard, as the approach requires us to formally codify our theoretical knowledge of a system in the form of an explicit computational model of the processes underlying it.  Through simulations we can produce counterfactuals, allowing us to evaluate which contextual variables we may exclude as conditioning variables, and whether the stability assumption is tenable.  If the stability assumption does not hold, we can examine the effects of the conditioning variables on the relationship of interest.  All the while we are able to model system processes explicitly, including non-linearities and feedback loops.

%When data is scarce, or sparse, as it is often the case with human complex systems, the statistical approaches become rather helpless, but properly specified theoretical causal model can still be usefully used to increase our understanding of the system's behaviour, through sensitivity analysis and the simulation of `what-if' scenarios. Indeed, we claim that ABM may be a very effective methodology to develop such theoretical models when investigating complex systems. With ABM, we formally codify our theoretical knowledge of the system in a computational model which produces (through simulations) counterfactuals, allowing us to assess which contextual variables we may \textit{not} consider in the list of conditioning variables, the extent to which the stability assumption is tenable and, if not, which kind of effects on the relationship of interest we may expect from the conditioning variables. 

%In this regard, rather than as a substitute of the traditional approach \textit{tout court}, ABM can be more properly seen as a complementary tool to assess the limits of the statistical approach when it deals with variables which are part of complex systems and, more generally, to investigate the behaviour of these systems in situations when the scarcity of suitable quantitative data warns us against the adoption of statistical approaches.

In this context, ABM can be seen as a complementary tool to assess the limits of statistical approaches as applied to a complex system, and to investigate system behaviour when quantitative data is too scarce to perform robust statistical analyses.

\section{A new paradigm for complex public health}

While ABMs offer great promise for the population health community, we suggest that their utility is too often evaluated from the viewpoint of mainstream traditional epidemiological methods \cite{marshall2015,murray2017}.  Moreover, some population health models described as `agent-based' would be better categorised as `individual-based', as individuals are represented but display little or no capacity for agency or interaction \cite{cherng2016,schaefer2013}. These issues have resulted in ABM studies being isolated experiments within the discipline, as their strengths are not being properly utilised, rather than developing into a mature category of population health research in their own right.  

However, as explained above, ABMs can have important and distinct uses in public health, exploring research questions that traditional methods cannot usefully address. 
ABMs allow us to understand how the combination of agent interactions and non-linear causal effects gives rise to system-level patterns. Such an approach falls outside the traditional epidemiological toolkit, but developing our understanding of these patterns is essential for improving the health of the public. With this in mind we would concur with Auchincloss and Diez-Roux, who outlined how regression approaches can limit our enquiries:

\begin{quote}
Too often, the exclusive use of regression approaches constrains not only the answers we get but also the types of questions we pose and the hypotheses and even theoretical explanations that we develop. In our search for what is `tractable' in empirical observational research (essentially that which mimics the perfect experiment), our questions have the tendency to become narrower and narrower and perhaps less relevant to understanding or intervening in the real world \cite[p. 6]{auchincloss2008}.
\end{quote}

Long after this paper was first published, the methodological debate in population health continues to discuss the relative merits of ABMs compared to regression in the presence of interference or dynamic processes \cite{naimi2016}. In our view, the debate should acknowledge that the primary advantage of a model-centred approach is that it allows us to answer fundamentally \emph{different} types of questions. Consider the ability of ABMs to integrate variables of disparate nature, including qualitative, narrative, and anecdotal \cite{yang2008}; to enable the inclusion of geographic and social space at multiple levels of resolution -- embedding feature-rich GIS representations, for instance, or sophisticated social network structures; and to simulate multiple complex processes simultaneously, all while remaining tractable (to a certain extent).

% Think of the frequently observed association between neighborhood poverty and rates of teenage pregnancy. A superficial interpretation of this process would be one that translates this aggregate correlation into individuals’ decisions: as poverty increases, the likelihood of a young woman having a child goes up by some amount. [ref.]

%An ABM based approach, in contrast, would explore the phenomenon taking into account at least a portion of the complex network of underlying motivations for these women’s decisions: how other people may influence those decisions (e.g. parents, partners, peers), how decisions may be mediated by women’s beliefs about the benefits and costs of having a child, by the type of support available in the community or from institutions, by the number of other teenagers becoming pregnant in a woman's friendship network or area of residence. Inputs for a model willing to encompass such considerations could come from administrative data, as well as ethnographic accounts of women's past experiences, qualitative interviews in the local or similar areas, longitudinal studies, etc.

%Allowing to take into account the underlying motivations for behavior, the feedback loops between the individual and the wider social and institutional environment, ABMs can support policy analysis, offering the ability to make inferences about how women would, for example, behave under alternative scenarios.

These features enable the exploration of processes in a fundamentally different way, and the exploration of fundamentally different questions of wider scope and breadth. Simply put, adopting ABMs requires not just a change of methods, but a change of mindset and conceptual framework.

\section{Towards Model-Based Science}

The radical change of mindset needed in order for agent-based modelling to gain traction in population health is one of the reasons why the uptake of ABMs in the discipline is currently minuscule (systematic reviews dig up mere tens of papers, despite ABMs having been used to study human social systems more generally for nearly 50 years \cite{nianogo2015}), and well below the prominence of other methods such as multilevel modelling.  This factor alone is not enough to explain the lack of engagement with ABM, however; other systemic and practical reasons contribute to the current state of affairs.

We suggest there are three main factors driving this lack of enthusiasm for the approach. Firstly, claims of what ABM can offer above traditional methods are contested. Newer statistical models may be capable of estimating causal effects in the presence of dynamic processes, treatment interference or spatial and network autocorrelation \cite{naimi2016}, thus removing some justifications for choosing ABMs over an alternative approach with which the community is already familiar. 

Secondly, computer programming\footnote{We refer to computer programming in the sense of using general-purpose programming languages, like C/C++, Python or Java, to construct self-contained computer programs.  This is a distinct skill set from using domain-specific scripting languages, like R or Stata, to construct statistical models.}, simulation modelling and complexity science are relatively uncommon skills in population health practice and research -- unlike statistical theory, data management, and quantitative analysis.  This is due to the traditional epidemiological focus on Positivist hypothesis-testing approaches to identify causal processes and intervention effectiveness.  Adoption of ABMs, in contrast to novel statistical techniques which build on existing conceptual frameworks and skill sets, may require substantial retraining and changes in the way in which people think about and characterise population health challenges.

Finally, as mentioned above, there is a third crucial factor: ABMs within population health are frequently compared like-for-like with approaches designed to answer fundamentally different research questions. As a consequence, the methodology is perceived as imprecise and intractable in comparison to statistical methods, when in reality the ABM is simply better-suited for different types of questions.

In order to address these factors, population health researchers will need more exposure to ABMs applied to relevant problems, more opportunities to develop the skills needed to build simulations, and will need to apply their models to appropriate problems that leverage the strengths of ABMs while managing their weaknesses.  We propose that a realignment of the discipline's approach to data and modelling methodologies more generally will be needed, in order for ABMs to sit comfortably amongst the other methods in population health's toolbox.   

%One discipline that in recent years has successfully transitioned from a purely statistical towards an increasingly model-based approach is demography. The evolution of modelling methods in the population sciences is a good example of how theory-driven, model-based approaches can alter a discipline and its approach to data.

\subsection{The conceptual challenges of model-based science}

Similar to other statistically-focused and empirically-inclined disciplines like demography \cite{courgeau2017}, population health has progressed methodologically in a cumulative fashion.  New statistical methodologies have spurred significant changes in research practices, yet each new methodology is fundamentally related to the previous, and no new method has completely replaced older ones.  Each statistical tool still has its place in the metaphorical population health toolbox.

%Courgeau et al. (2017) take stock of the progress of demography over the centuries, and propose that the discipline is on the cusp of a new methodological paradigm.  They note that the progress of demography to date has been a cumulative journey -- each new methodological revolution has enhanced the previous, and yet no new methodology has completely replaced any previous one.  Each methodological tool still has its place in the metaphorical toolbox. 

Agent-based modelling, however, does not fit this pattern as cleanly. ABMs are not a statistical methodology, and thus cannot simply be added to the methodological toolbox and applied to the same problems in the same way.  If we take up multi-level modelling, for example, previous data can still be used and the methods themselves provide some guidance about what new data may prove useful. ABMs, however, have a more complex relationship with data, and in general what keeps the modeller awake at night is not a lack of data but instead the need for sensible parameter values for specifying agent behaviour.

So the newly-minted ABM researcher in population health must think not only about data sources, but some new topics as well:
\begin{itemize}
    \item Social and behavioural theories: what should our agents be doing, and why?
    \item Uses beyond prediction: what can the ABM add to our knowledge of population change?
    \item Seeking out more diverse forms of data: can we alleviate our parameter problems using a wider variety of data sources, like qualitative data, GIS information, or participatory group model building?
\end{itemize}

This kind of shift in practice requires not only knowledge of a new approach and its epistemological and computational limitations, but a new perspective on how and when the approach should be used, and where it should fit in the overall scientific spectrum of population health.

Given the significant epistemological and practical implications of adopting agent-based methods, we propose that the relationship between ABMs and population health needs to change in a substantive fashion before their usefulness becomes apparent.  The development of a theory-based, model-centric approach to certain problems is needed, and in so doing we will uncover areas where ABMs are more suitable.

In our view, ABMs would best serve the interests of population health by being applied to areas where individual behaviour and interactions between individuals and their environments are influential in determining the success of interventions.  In the next sections, we will discuss in more detail the potential role ABMs can play in a model-centric vision of population health, present some methods to mitigate the weaknesses of the approach, and finally illustrate some key areas of research where we believe ABMs can make a strong contribution.
%we outline how ABMs may be able to provide new and valuable insight into the `wicked problems' facing population health.

%In demography, these problems have been focused on explanatory aims, and most commonly applied to areas where simply `throwing data at the problem' is not proving successful.  Migration, for example, is commonly seen as the most uncertain aspect of population models in demography, so ABMs have been applied to migration to examine possible motives for migration behaviours in various contexts (see e.g., Willekens 2012, Klabunde 2014).  Similarly, in the context of public health ABMs can be applied particularly well to areas where sociality is influential in determining the success of interventions. 

\section{Roles for Simulation in Population Health}

Some of the most challenging `wicked' problems in population health involve complex, interacting processes that are difficult to characterise in a traditional epidemiological framework.  These challenges are likely to benefit from ABM research, given that they can model the social and environmental aspects of population health explicitly.

However, this still begs the question: what would these models actually do?  How would our quest to reduce or eliminate a `wicked' problem benefit from an ABM, and how would it add to our knowledge?  Here we illustrate a few ways in which ABM approaches can enhance our efforts.

\subsection{Models as Policy Sandboxes}

%Evidence for population health policy is often based on causal inference derived from health technology assessment methods. However, health policy-making frequently considers factors other than evidence on causes. Given the evidence of poor transferability of interventions, looking to other factors seems a sensible approach, as much as it causes difficulties for demonstrating research impact.  

ABMs enable us to build a policy sandbox; a place to explore what evidence would be useful for taking certain decisions; to develop collaboratively a theory for why a certain policy will or will not work; and to test out how interventions could have an effect, if these theories hold true. Di Paolo, Noble and Bullock propose that while ABMs are often approached as a method to develop `realistic simulacra' of the physical world, they may be best placed as platforms for the exploration of theoretical relationships within a system, their interactions and consequences \cite{dipaolo2000}. Such conceptual exploration can be very useful for building theory. An approach that facilitates the apprehension of concepts in an abstract, rather than data-driven sense is useful for the precise reason that it differs from the data-driven approach of orthodox epidemiology. 

In essence, traditional epidemiology focusses on risk factors and outcomes, with the link between them being an opaque black box \cite{susser1996}. However, in order to develop interventions to change the relationship between risks and outcomes, what is within that black box itself needs to be changed. This requires understanding mechanisms and having a theory of change in order to modify them. Statistical methodology does not provide the tools to model mechanisms, so researchers tend to focus on tasks that statistical methods can solve, which is the description of inequalities identified by the data without necessarily providing solutions. ABMs, in contrast, require people to model mechanisms explicitly, at least in the abstract, and provide a set of tools which encourages people to focus on the parts of the system where change might occur and could have the greatest impact. This leads us more toward solutions-focused research.

As Marshall notes, ABMs also provide a means to make use of a wider range of evidence: 

\begin{quote}
Agent-based modeling represents one (but not the only) method to synthesize prior knowledge of a population -- and the causal structures that act on this population -- to understand how an intervention could affect the public's health. In this manner, agent-based modeling is a science of evidence synthesis. Specifically, ABMs (and other simulation approaches) represent a platform for the integration of diverse evidence sources, including inconsistent or inconclusive scientific information, to support decision making for complex public health problems. \cite{marshall2017}
\end{quote}

Further, the ability of ABMs to serve as theoretical exploration tools and intuition pumps when data is sparse can prove to be a significant advantage \cite{silverman2011}, as modellers would be able to investigate theories about population health using ABMs even in the absence of the expensive, difficult-to-collect data upon which statistical approaches depend.  Engaging in this kind of `model-based science' can enable the evaluation of complex policy interventions amongst simulated populations, with outcomes serving as guides to decision-making under complexity rather than as point predictions of intervention outcomes. 
%The rise of ABM has even been presented as the appearance of an altogether new approach to demography as a model-based science -- one in which the study of rich interactions within a population form a central part of the understanding of population trends at the higher level \cite{corgeau07, courgeau2017}.

There are additional pragmatic benefits to data-light, model-based investigations of interventions.  Health data can be expensive to collect, as well as legally challenging due to strict data protection laws.  Being able to circumvent these problems and use generative approaches to understand complex health problems would allow for potentially fruitful investigations of potential policy interventions when data is hard to obtain.  Further, the outcomes of our simulations may help us identify where future data collection is needed.  

\subsection{Investigating Assumptions}

ABMs have provided a method to run in silico interventions to inform policy makers about options, based on counterfactual model scenarios featuring agent populations using the same decision rules under varied environmental conditions.  This scenario-based approach could be further augmented by allowing policy-makers direct input into which elements of proposed interventions are implemented, and the behavioural mechanisms underlying agent behaviours.
%They could go further, and allow policy makers direct input into which elements of intervention theory are more or less plausible in their local context, to modify the generative mechanisms that underpin the behaviour of agent in the system and change the specification of a causal theory. 

Traditional analysis of data is agnostic to the audience. ABMs in contrast provide a means to conduct post-normal science, or a scientific enterprise conducted in a context where `the puzzle-solving exercises of normal science (in the Kuhnian sense)  which were so successfully extended from the laboratory to the conquest of Nature, are no longer appropriate for the resolution of policy issues of risk and the environment' \cite[p. 750]{funtowicz1993}.  Post-normal science acknowledges the pivotal role of input from the public and from policy makers when dealing with environments suffused with unavoidable uncertainty.  In this context, generative theories of causation still privilege scientific knowledge, but also integrate and adapt to the expertise of knowledge-users.  The post-normal paradigm allows evidence to inform policies to improve health, moving beyond a context-free evidence approach.
%In post-normal science, decisions with huge stakes must be taken in an environment where the scientific knowledge available is unavoidably uncertain, and as a consequence input from the public and from policy-makers becomes critical. Generative theories of causation appropriately privilege the body of scientific knowledge, but also integrate with, and adapt in response to the expertise of knowledge users. This provides a paradigm where evidence can inform policies to improve health, moving beyond a context-free evidence approach. 

Post-normal science does not yield to the opinions, values or politics of knowledge-users, and still adheres to the principles of calibration, validation and verification.  The post-normal ABM approach provides opportunities for direct comparison of scenarios that include broader knowledge with those based solely on empirical data collection.  Future research can examine how these approaches to model-building compare in their ability to inform health policy work and produce effective interventions.

%This does not mean that the science yields to opinions, values or political will of knowledge users - all the principles of verification, validation and calibration are still integral, but now the ABM provides direct comparison on the validity of \emph{in silico} systems between those that do and do not incorporate broader knowledge. A fruitful avenue for future research is studying the extent to which human-calibration versus data-calibration produces models that can usefully inform health policy work and produce effective policies.

From the perspective of post-normal science, ABMs function as a mechanism for reaching a collaborative agreement with knowledge-users about the emergence of population-level patterns in health from low-level interactions in social and spatial environments.  The model provides feedback on the validity of the shared assumptions underpinning our understanding of critical population health issues.  Exploring the model's parameter space or implementing differing behavioural mechanisms enable us to compare different assumptions.  In this paradigm, the process of model-building itself is an avenue for generating, developing, and comparing knowledge, and in the process, informing empirical and practical work on the phenomenon being modelled.

%Rather than setting our minds to the task of how ABMs can better draw on the evidence base,  ABMs provide a mechanism of reaching a collaborative agreement (drawing upon various forms of evidence and input) over how interactions in social and spatial environments give rise to population-level patterns. An ABM provides feedback on the validity of the shared assumptions underpinning a population health issue. Differing assumptions between research evidence and evidence users can be investigated by exploring an ABM's parameter space, or implementing differing mechanisms for behaviours that instantiate different assumptions about the underlying mechanisms.

\section{Understanding ABMs}

As we have outlined, ABMs have numerous advantages when one wishes to explore the behaviour of a complex system.  However, as with any methodology, ABMs are not a panacea and have aspects which can present significant challenges for the modeller.

One area in which ABMs remain opaque in some respects is in the analysis of their results.  ABMs by their nature are suitable for modelling the behaviour and evolution of systems which defy formal statistical analysis, and that very complexity means they can appear opaque, where inputs and outputs are present but the influence of processes within the model are tangled and unclear.  As a consequence, finding suitable parameter values for a simulation can be a lengthy process of trial and error, leading to significant time investment and frustration for the modellers.  

This is an area where cutting-edge advancements in statistical and machine learning can work in concert with simulation modelling to produce insightful results.  The growth of the field of uncertainty quantification and the current accessibility of machine learning opens up new possibilities for making ABMs more transparent and reducing these barriers to wider adoption.

\subsection{Uncertainty quantification}

The advent of the Bayesian Analysis of Computer Codes (BACCO) methodology has produced significant advancement in the analysis of model uncertainty and the impact of parameter values \cite{ohagan06}.  In particular, Gaussian Process Emulators (GPEs) have proven to be a promising method for analysing ABMs.  In essence, a GPE takes a training set consisting of simulation outputs resulting from a wide range of input parameter values taken from across the parameter space, then develops a `surrogate model' -- a statistical model of the original simulation model \cite{kennedy01}.  The result is a summary of the fraction of the final output variance accounted for by each input parameter, as well as their interactions.  This gives us a much clearer picture of the impact of each input parameter.  
%GPEs assume that the output variable of interest can be decomposed into main effects derived from input parameter values, alongside a constant term, and the uncertainty introduced by the computer code itself (e.g., rounding errors).  

GPEs have been applied successfully in ABMs of various types, including models of social care \cite{silverman2013soc}, research funding allocation \cite{silverman2016}, and the effect of landscape changes on bird populations \cite{parry2013}.  These analyses vastly increase the interpretability of ABM results, and provide much clearer guidance as to the function of each simulation parameter.  Much of this work has been facilitated by the availability of GEM-SA, a free software package for analysing simulation results using GPEs \cite{kennedy2004}.  As this technique continues to mature, and additional user-friendly means of implementation are created, variations of the GPE approach seem likely to become a common method for better understanding the internals of an ABM.

%GPEs and related techniques can also be applied before the ABM parameter settings are finalised.  If particular parameter values are uncertain, or their values are unknown or difficult to estimate, modellers can use GPEs to develop a training set covering a wide portion of the parameter space, and then use the resultant sensitivity analysis to identify portions of the space that produce realistic values for the main output values of interest.  These areas can then be examined in greater detail with focused series of runs of the full simulation.  Thus, the GPEs can help identify reasonable values for unknown parameters, and reduce the computational expense of doing so by running the emulated model rather than the full simulation.

\subsection{Machine-learning surrogate models}
 
%For instance, machine learning has been proposed in combination with ABM to achieve automatic calibration of simple trajectory samples, where agent-generated movement is used as an input of a machine-learning pipeline \cite{torrens2011building}. 

Machine learning, similar to GPEs, can also be used to develop `surrogate models' which can allow for a detailed examination of the model's behaviour and parameter space without an explosion of computational demands \cite{pereda2017brief,lamperti2017agent}.  Machine learning methods can improve the theoretical understanding of the ABM, help calibrate the model, and facilitate interpretation of the results relevant to end-users, therefore achieving an overall model of complex health issues with greater empirical validity.   

Deep neural networks can also be used to create surrogate models, by training a multi-layered neural network on the output of numerous simulation runs \cite{van2017deep}. Neural networks are able to approximate the output of any function, meaning that a trained neural network can approximate an ABM's output to a high degree of precision with much lower computational requirements.  Neural-network surrogate models thus allow for much easier probing of an ABM's behaviour, and enable detailed sensitivity analyses to be performed.  The surrogate model can also point us toward interesting areas of the parameter space that merit further exploration with the full simulation. 

We should note that neural networks notoriously suffer from the issue of low interpretability, due to their numerous parameters, the complex structure of multiple hidden layers, and inherent non-linearity. However, we argue that interpretability is not a key requirement when using them as a surrogate of an ABM. If explaining the simulation result is of importance to the user, interpreting the ABM directly would be more suitable than explaining the prediction of the surrogate model.  The trained neural network is intended to replicate the behaviour of the ABM to allow us to generate analyses of the model's behaviour more quickly.  In this pragmatic role, our primary concern is that the neural network provides significant computational savings, not the particular way in which the network learns to replicate the ABM.
%This is due to the fact that the trained neural network surrogate will be optimised to mimic the ABM results, but might not base its structure on the actual structure of the ABM, even when the `surrogate vs ABM' error is negligible.

With the advent of these sophisticated methods for investigating the outputs of ABMs, we can begin to reduce their opacity and use them more effectively to understand how complex systems operate.  Using surrogate models enables us to reduce an ABM's computational demands, while also facilitating deeper analyses of the model's underlying mechanisms.  These analyses can then help us to understand how we can most effectively intervene at the micro-level of a population to produce positive changes at the macro-level.

\section{Ways forward}

As we have seen, ABMs and concepts drawn from the study of complexity could have a valuable role to play in future research in population health.  The `wicked' problems of the 21st century present a significant challenge to traditional statistical approaches, so approaching them from a complex systems perspective can enable us to characterise these problems in terms of the social and environmental interactions which lie at their core.

As with any field, not every tool is fit for every task, and this remains the case with ABMs and related methods.  In the case of less `wicked' issues in population health where causal links are more clearly defined, statistical methods already available and easily implemented can work quite well to generate the kind of effect estimates health researchers might seek.  ABM should be a significant and powerful addition to the toolbox for population health modellers, used in harmony with the trusted epidemiological methods already in evidence.   

For efforts extending further into explanation and description of more complex problems, ABMs start to shine.  Below we describe three areas of potential interest population health researchers keen to experiment with ABMs, and note the difficulties facing traditional approaches in these contexts.

\subsection{Health Inequalities}

Health inequalities, which are systematic differences in the health status of different population groups, remain a difficult problem even in countries with otherwise high levels of equality \cite{bambra2011}. For example, Nordic countries, despite generous welfare states and generally excellent population health statistics, still have persistent health inequalities \cite{shaw2014}. Likewise, health inequalities have persisted in the United Kingdom even in the presence of free health care and systematic attempts by the New Labour government to reduce them \cite{mackenbach2011}.  

A potential reason for this persistence is the sheer level of social reform that would be required to tackle this problem \cite{mackenbach2011}.  Wealth inequalities might need to be corrected, access to healthcare and hospital/clinic locations might need to be changed and optimised, and significant educational efforts to encourage changed health behaviours might need to be undertaken, amongst other changes.  In the real world, policy interventions on this scale are not only cost-prohibitive, but quite disconcerting to policy makers, who are (perhaps understandably) reticent to restructure the whole of society on the basis of theories which in effect cannot be implemented without tremendous risk of unintended consequences \cite{berkman2011}.  These unintended consequences, or spillover effects, can potentially impact negatively on other key areas of population health \cite{lorenc2014}. 

ABMs, however, provide a testbed of sorts.  Generative approaches and virtual populations can provide a risk-free laboratory for implementing large-scale policy interventions on simulated populations.  While we would not wish to claim these kinds of models necessarily provide strong predictive power, if given sufficient detail in relevant social and behavioural mechanisms they could provide important insights on potential unintended consequences of these large-scale policy interventions.

More broadly, ABMs can shed light on those areas where health effects at the population level appear to be at odds with our expectations, given our knowledge of individual-level behaviours and processes.  The case of inequality in Scandinavian welfare states above is a good example: we would expect very equal societies with universal access to free healthcare to show minimal health inequalities at the population level but, in reality, health inequalities are both still present and surprisingly significant.  ABMs and complex systems approaches could help population health researchers unravel these mysteries by modelling the social and environmental effects that might lead to differential health outcomes despite society being set up specifically to avoid those outcomes.

\subsection{Alcohol Use and Misuse}

In the UK, the harm caused by alcohol both to the individual and to others is significantly higher than that caused by other drugs \cite{nutt2010}. The role of psychological \cite{hammersley2014}, social \cite{meier2018} and environmental factors \cite{birckmayer2004} in influencing levels of substance use and dependence are increasingly recognised as key factors beyond biological addiction processes, both in terms of the risk processes, and for the most appropriate points of intervention. Scotland has recently implemented a minimum price per unit of alcohol \cite{brennan2014}. 

Predicting the outcomes of population-level interventions is far from straightforward. Price-based measures are more likely to be effective amongst drinkers on the lower end of the socioeconomic scale, whereas changes in pricing will have a greater impact on day-to-day spending habits \cite{holmes2014}, but availability of alcohol (at any price) is strongly patterned by geography, with deprived areas of Glasgow having a greater density of alcohol, tobacco, fast food and gambling outlets, for example \cite{macdonald2018}. How alcohol retailers respond to minimum price legislation in areas of high compared to low supply may influence, and be influenced by changes in purchasing behaviour. Traditional statistical approaches are limited in the extent to which they can study non-linear interactions such as those between individuals and local retailers.   

ABMs and complexity-inspired approaches can make substantive contributions in this area.  Statistical modelling studies have looked in-depth at minimum unit pricing policies, but computational models could allow us to investigate how interactions between individuals, individuals and their neighbourhoods, and  related effects such as socioeconomic differences \cite{holmes2014}.  Alcohol and drug abuse treatment outcomes are also affected by social factors \cite{schroeder2001,homish2008}, and modelling these interactions and networks via ABMs could help in further fine-tuning interventions in this respect \cite{yorghos2018}

\subsection{Obesity}

Obesity is the target of substantial research and investment in population health worldwide, given that the condition is linked to such a wide range of health problems -- both as an effect and a cause.  Obesity stands out amongst the target research areas listed here, in that a number of simulation studies have already targeted obesity, and there have been some high-profile calls for systems-based modelling of the problem.  The Institute of Medicine presented a book-length study of obesity prevention efforts in the United States, concluding that researchers and policy makers must take a `systems thinking' perspective \cite{IOM2010}.  Chapter 4 provides a useful look at systems perspectives on the obesity epidemic -- useful background for interested modellers \cite{IOM2010}.  Similarly, Foresight in the UK was commissioned to develop a strategic 40-year plan to tackle obesity in the UK, and presented a whole-system approach to the problem advocating large-scale societal, personal, and governmental changes \cite{Foresight2007}.

Skinner et al. outline nine properties which make obesity well-suited as a target for systems-science-based modelling efforts, and present a systematic review of systems science models in obesity research (\cite{skinner2013}).   While their review finds dozens of studies using systems science perspectives and techniques, they conclude that the work done thus far has taken relatively limited views of the topic, investigating manageable portions of the overall obesity problem rather than tackling the larger, messier complex of societal and individual factors which drive obesity trends.  Levy et al. take a similar view, noting that obesity modelling efforts are `at a nascent stage of development' and just focus on `one or two links in the process', leading the authors to advocate a comparative modelling approach \cite[p. 390]{levy2011}. 

Obesity is thus a highly relevant area for modellers of a complex systems background, and indeed work of this type is already underway \cite{silverman2017}.  Much of the conceptual work around the social, behavioural and environmental factors influencing the obesity epidemic has already been done by respected sources.  The widespread agreement that ambitious societal change at multiple levels will be required clears the way for similarly ambitious modelling projects to attempt to understand the interplay of the multitudinous factors at play in obesity, and to develop productive collaborations with population health researchers and governmental bodies.

\section{Conclusion}

While population health research has contributed to numerous high-profile health successes in modern times, there remain some highly complex, `wicked' problems which defy traditional methods of epidemiological analyses, and have resisted our attempts to develop effective interventions at the population level.  These wicked issues in population health fall into the category of post-normal science -- `These issues are urgent and of high public and political concern; the people involved hold strong positions based on their values, and the science is complex, incomplete and uncertain' \cite[p. 163]{gluckman2014}. Whereas traditional epidemiological methods are designed to reduce uncertainty about specific causal relationships, reliance purely on these methods inhibits the ability of the population health research community to say anything about complex wicked issues that urgently require solutions and where the scientific voice is non-existent or overwhelmed by caveats and uncertainty. Methods designed to illuminate complex interrelationships are unlikely to provide certain answers, but can provide an important contribution to the debate and may also provide a mechanism to bring together a range of alternative sources of knowledge. 

As outlined above, there are numerous areas in which a complexity-inspired approach could contribute to efforts to develop more robust population health interventions for these wicked issues.  These very brief examples demonstrate that developing interventions in these areas will require a broader, systems-based perspective due to the significant societal and behavioural change required.  The complex systems science community is notable for its ability to apply computational approaches to an enormous range of domains, from artificial chemistry to simulated societies, by modelling the unifying mechanics and principles that describe the complex behaviour of these seemingly disparate systems.  We suggest that this perspective makes this approach uniquely well-suited to undertake ambitious and challenging modelling projects aimed squarely at these wicked health problems.

Such efforts would require significant investment, not just in terms of time and finances, but in the form of interdisciplinary communication, which is no simple undertaking.  Developing a common language between population health and complex systems will take time, but the potential benefit to efforts to enhance the health and wellbeing of millions of people seem worth the risk of occasional frustration and misunderstanding.

\begin{backmatter}
\section*{Competing interests}
  The authors declare that they have no competing interests.
\section*{Funding}
This work was supported by UK Prevention Research Partnership MR/S037594/1, which is funded by the British Heart Foundation, Cancer Research UK, Chief Scientist Office of the Scottish Government Health and Social Care Directorates, Engineering and Physical Sciences Research Council, Economic and Social Research Council, Health and Social Care Research and Development Division (Welsh Government), Medical Research Council, National Institute for Health Research, Natural Environment Research Council, Public Health Agency (Northern Ireland), The Health Foundation and Wellcome.  

MMcC holds a Medical Research Council/University fellowship supported by MRC partnership grant (MC/PC/13 027). ES are in the Complexity in Health Improvement Programme supported by the Medical Research Council (MC\_UU\_12017/14) and the Chief Scientist Office (SPHSU14). MMcC is in the Relationships programme supported by Medical Research Council (MC\_UU\_12017/11) and Chief Scientist Office (SPHSU11). RS is supported by a MRC Mental Health Data Pathfinder Award (MC\_PC\_17217). 

These funders provided financial support for the authors, but were not involved in the development of this work or the preparation of this manuscript.

\section*{Author's contributions}
    ES, RS, and MM developed the concept for this paper and wrote the original drafts of the manuscript.  UG, SP and JA significantly expanded the sections relating to complexity science core concepts.  CA contributed the section on machine-learning surrogate models.  All authors reviewed and edited the final manuscript in collaboration.

\section*{Acknowledgments}
CA would like to acknowledge the support of Research England's THYME Project.

\bibliography{BMC_2020_Silverman}

%\begin{table}
%%\hyphenpenalty10000
%\centering
% \begin{tabular}{| p{2.3cm} | p{2.3cm} | p{2.3cm} | p{2.3cm} |} 
% \hline
% \textbf{Research\newline Question} & \textbf{Policy\newline Decisions} & %\textbf{Research\newline Method} & \textbf{Research\newline Input} \\ [0.5ex] 
% \hline\hline
%Does device X reduce health outcome Y? & Provide or prevent access to device X & %Randomised trial allocating X or no X & Participant data, device utilisation \\ 
% \hline
%Does device X reduce health outcome Y?\newline(Impossible to conduct RCT) & Provide or %prevent access to device X & Cohort study of device X use\newline(Causal inference %methods) & Participant data, device utilisation, recontact and follow-up \\
% \hline
%Will people use device X? & Provide or prevent access to device X & Qualitative %interviews & Participant interviews, transcription analysis and interpretation \\
% \hline
%What influences (or could influence) population patterns in device X use? & Identify conditions of provision or prevention in local context & Agent-based Models & Policymaker discussion, theory, quant data, interview data \\ 
% \hline
%\end{tabular}
%\caption{Comparing research methods, inputs and outcomes for differing research questions}
%\label{table:1}
%\end{table}
\end{backmatter}
\end{document}